\documentclass[journal=jacsat,manuscript=article]{achemso}

\usepackage[version=3]{mhchem} 
\usepackage{xcolor}
\usepackage{booktabs}
\usepackage[numbers]{natbib}  
\usepackage{xcolor}            
\usepackage{graphicx}          
\usepackage{float}
\usepackage{subcaption}

\usepackage{graphicx,lipsum}  
\usepackage{subcaption,xcolor}  

\author{Anushree Dutta}
\affiliation{School of Chemical Sciences, Indian Association for the Cultivation of Science, Kolkata-700032, India}

\author{Pintu Mandal}
\email{pintuphys@gmail.com}
\affiliation{Department of Physics, St. Paul's Cathedral Mission College, Kolkata-700009, India}

\author{Nabanita Deb}
\email{nabanita.deb@iacs.res.in}
\affiliation{School of Chemical Sciences, Indian Association for the Cultivation of Science, Kolkata-700032, India}

\title[An \textsf{achemso} demo]
 {Influence of octupole field on quadrupole mass filter performance in the second stability zone}

\abbreviations{IR,NMR,UV}

\begin{document}

\begin{abstract}
Radial asymmetry in a quadrupole mass filter (QMF), introduced by symmetric displacement of a diagonally opposite rod pair, modifies the confinement potential and alters the ion stability characteristics. In this work, the influence of such radial asymmetry on QMF operation in the second stability zone is investigated through simulations. Using an existing potential formulation for a radially asymmetric QMF, the stability diagram in the second stability zone is extracted for the first time, revealing a systematic shift of the stability apex with the asymmetry parameter. Radial asymmetry introduces additional multipole components, notably an octupole term whose magnitude scales with the asymmetry parameter and whose sign depends on the DC polarity applied to the displaced rods. Transmission simulations show that the transmission peak shifts in accordance with the displaced stability apex, while the resolution exhibits a strong dependence on both the asymmetry parameter and the DC polarity. Comparable maximum resolution is obtained for inward and outward displacements under suitable polarity configurations, with outward displacement providing higher transmission efficiency due to an increased effective aperture. These results demonstrate that controlled radial asymmetry provides an effective means to tailor the resolution and transmission characteristics of QMFs operating in the second stability zone.

\end{abstract}

\textbf{keywords}: Quadrupole mass filter, radial asymmetry, stability diagram, octupole field, transmission characteristics, resolution, RF phase selectivity

\section{Introduction} \label{sec1}

Quadrupole mass filters (QMFs) are among the most established mass analyzers, with ion stability governed by the Mathieu equations arising from a time-dependent quadrupolar electric field~\cite{paul1953neues,paul1955elektrische, miller1986quadrupole}. Since the earliest demonstrations of quadrupole mass filtering, stable ion transmission has been associated with discrete regions in the Mathieu stability diagram. Although multiple stability zones exist, practical QMF operation has traditionally been confined to the first stability zone~\cite{li2021towards,konenkov2002matrix}. Extensive theoretical, numerical, and experimental work has been carried out on the operation of quadrupole mass filters in the first stability zone\cite{march2005quadrupole, dawson2013quadrupole, douglas1992collisional}, which remains the standard operating regime due to its favorable balance between ion transmission and mass resolution. The continued pursuit of higher mass resolution has driven the exploration of QMF operations beyond the first stability zone. 
Operation of the QMF in the second stability zone imposes fundamental constraints due to its intrinsically narrow stability boundaries, which reduce ion acceptance and transmission efficiency. Ion motion in this zone is highly sensitive to small variations in RF amplitude and frequency, DC–RF ratio and ion energy. The requirement of higher RF voltages further enhances sensitivity to the instantaneous RF phase at injection, introducing strong phase selectivity in transmission \cite{holme1973dependence,ma1996simulation}. Despite these challenges, operation in the second stability zone has attracted sustained interest due to its potential for significantly enhanced mass resolution, improved mass spectral peak shapes including suppression of low-mass tailing. \cite{dawson1984second,konenkov1991characteristics,du1999peak,hogan2008performance} 

QMFs with symmetric circular-rod geometry inherently generate higher-order multipole field components, most notably dodecapole and icosapole terms, in addition to the ideal quadrupole field~\cite{hogan2009effects,schulte1999nonlinear}. Deviations from ideal symmetry arising from mechanical tolerances or intentionally introduced asymmetries, such as unequal electrode radii or electrode displacement, give rise to additional multipole components, including octupole and hexadecapole terms, which further influence QMF performance~\cite{JANA2025117495,jana2025radial,hogan2009effects,bugrov2023modelling}. In general, asymmetries introduced by the use of asymmetric rods or a diagonal pair of asymmetric rods have been shown to reduce mass resolution in the first stability zone due to the interplay between octupole and dodecapole components~\cite{JANA2025117495,jana2025radial}. In contrast, a controlled radial asymmetry introduced by displacing a diagonal pair of electrodes has been demonstrated to enhance the mass resolution~\cite{jana2025radial}. Ding et al. showed that introducing an octupole component in the range of 2–4\%, achieved using a diagonally opposite pair of smaller-radius electrodes biased with a positive DC potential, results in a substantial improvement in mass resolution for QMFs operated in the first stability zone~\cite{ding2003quadrupole}. Furthermore, Douglas et al. reported that QMF mass resolution strongly depends on the polarity of the octupole field~\cite{douglas2005linear}, and Moradian et al. demonstrated that higher ion detection efficiency can be achieved by adding a positive octupole component of approximately 2\%~\cite{moradian2007experimental}.

While the effects of mechanical tolerances on QMF performance have been investigated in the third stability zone by Hogan and Taylor~\cite{hogan2008performance}, Konenkov and co-workers theoretically analyzed nonlinear resonances arising from field imperfections and ion collection effects associated with the periodic ion motion in the second stability zone~\cite{du1999peak}. However, a systematic investigation of the impact of controlled radial asymmetry on QMF performance in the second stability zone has not yet been reported. 

In the present work, the second stability zone of radially asymmetric quadrupole mass filters is examined in detail, where an octupole field is introduced through controlled displacement of a diagonally opposite electrode pair. The corresponding stability diagrams and transmission contours are obtained using simulations, and the influence of both the magnitude and polarity of the octupole term on transmission characteristics and achievable mass resolution is systematically investigated.

\begin{figure}
    \centering
    \includegraphics[width=0.5\linewidth]{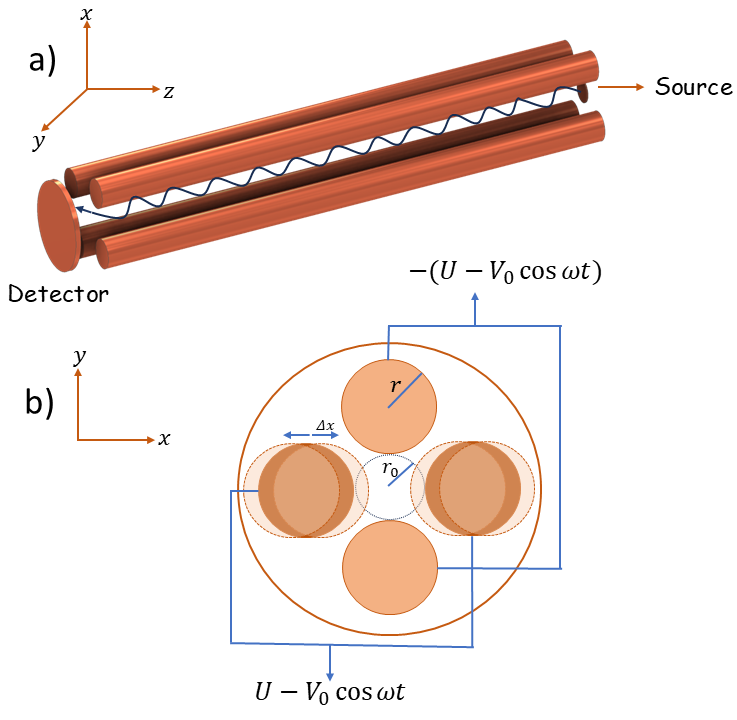}
    \caption{(a) Schematic of the linear QMF with circular rods. (b) Cross-sectional ($xy$ plane) view of the QMF with electrical connections, and illustrating diagonally opposite electrodes displaced from their normal positions.} 
    \vspace{-0.5cm}
    \label{fig:1}
\end{figure}

\section{Potential and Stability}
The quadrupole potential in a conventional symmetric round-rod mass filter is given by
\begin{equation}\label{eq1}
    \phi(x,y,t)=\frac{(x^2-y^2)}{r^2_{0}}\left(U-V_{0}\cos\omega t\right),
\end{equation}
where $r_0$ is the radius of the inscribed circle, as shown in figure~\ref{fig:1}. The ion dynamics in this time-dependent potential are governed by the Mathieu equation
\begin{equation}\label{eq2}
  \frac{d^2 u}{d\xi^2} + (a_u - 2q_u \cos 2\xi)u = 0,
\end{equation}
where $u=x,y$; $\xi=\omega t/2$ and
\begin{equation}\label{eq3}
    a_x=-a_y =\frac{8eU}{m r_{0}^2 \omega^2}, \quad q_x =-q_y = \frac{4eV_0}{m r_{0}^2 \omega^2}.
     \end{equation}
In the presence of radial asymmetry caused by radius variation or displacement of a diagonal pair of electrodes (figure.~\ref{fig:1}), the quadrupole potential deviates from eq.~(\ref{eq1}) and can be expressed as~\cite{jana2025radial}
\begin{equation}\label{eq4}
    \phi(x,y,t)=2\frac{(x^2-y^2)}{(x^2_{0}+y^2_{0})}\left(U-V_{0}\cos\omega t\right),
\end{equation}
where $2x_0$ and $2y_0$ are the surface-to-surface separations between the diagonally opposite pair of rods along the $x$ and $y$ directions, respectively, as shown in figure~\ref{fig:1}. For a symmetric displacement $\Delta x$ of a diagonal pair of electrodes along the $x$ direction, the effective electrode separations become $x_0=r_0 + \Delta x$ and $y_0 = r_0$. Substituting these into eq.~(\ref{eq4}), the potential can be rewritten as
\begin{equation}\label{eq5}
    \phi(x,y,t)=\frac{2(x^2-y^2)}{r^2_{0}\left(1+(1+\gamma_{x})^2\right)}\left(U-V_{0}\cos\omega t\right)=\frac{(x^2-y^2)}{r^2_{0}p}\left(U-V_{0}\cos\omega t\right),
\end{equation}
where $\gamma_x=\Delta x/r_0$ is the radial asymmetry factor and
\begin{equation}\label{eq6}
    p=\frac{1+\left(1+\gamma_x\right)^2}{2}
\end{equation}
is a geometric correction factor accounting for the radial asymmetry. The corresponding modified Mathieu equation takes the form
\begin{equation}\label{eq7}
  \frac{d^2 u}{d\xi^2} + \frac{1}{p}(a_u - 2q_u \cos 2\xi)u = 0.
\end{equation}
As evident from eqs.~(\ref{eq2}) and (\ref{eq7}), the Mathieu parameters $a_u$ and $q_u$ are effectively scaled by the factor $p$ in the presence of radial asymmetry. Consequently, the stability boundaries of the quadrupole mass filter are shifted relative to the symmetric case. In particular, the apex of the primary stability region in $q$-space is displaced to $q'=qp$, where $q=4eV_0/mr_{0}^{2}\omega^{2}$ is the apex of the second stability region of the symmetric setup. For small electrode displacements, $p \simeq 1 + \gamma_x$, which yields
\begin{equation}
q' \simeq q\left(1+\gamma_x\right).
\end{equation}

Here, $\gamma_x$ is positive for outward displacement of the rod pair and negative for inward displacement. Accordingly, the apex of the stability region shifts linearly toward higher $q$ values for outward displacements and toward lower $q$ values for inward displacements.

A linear shift in the apex of the first stability region in a radially asymmetric QMF has recently been reported by Jana et al.~\cite{jana2025radial}. In the following, the modification of the second stability region induced by such radial asymmetry is examined in detail using the Runge–Kutta 5(4) (RK45) method.

\begin{figure}[H]
    \centering
    \includegraphics[width=1\linewidth]{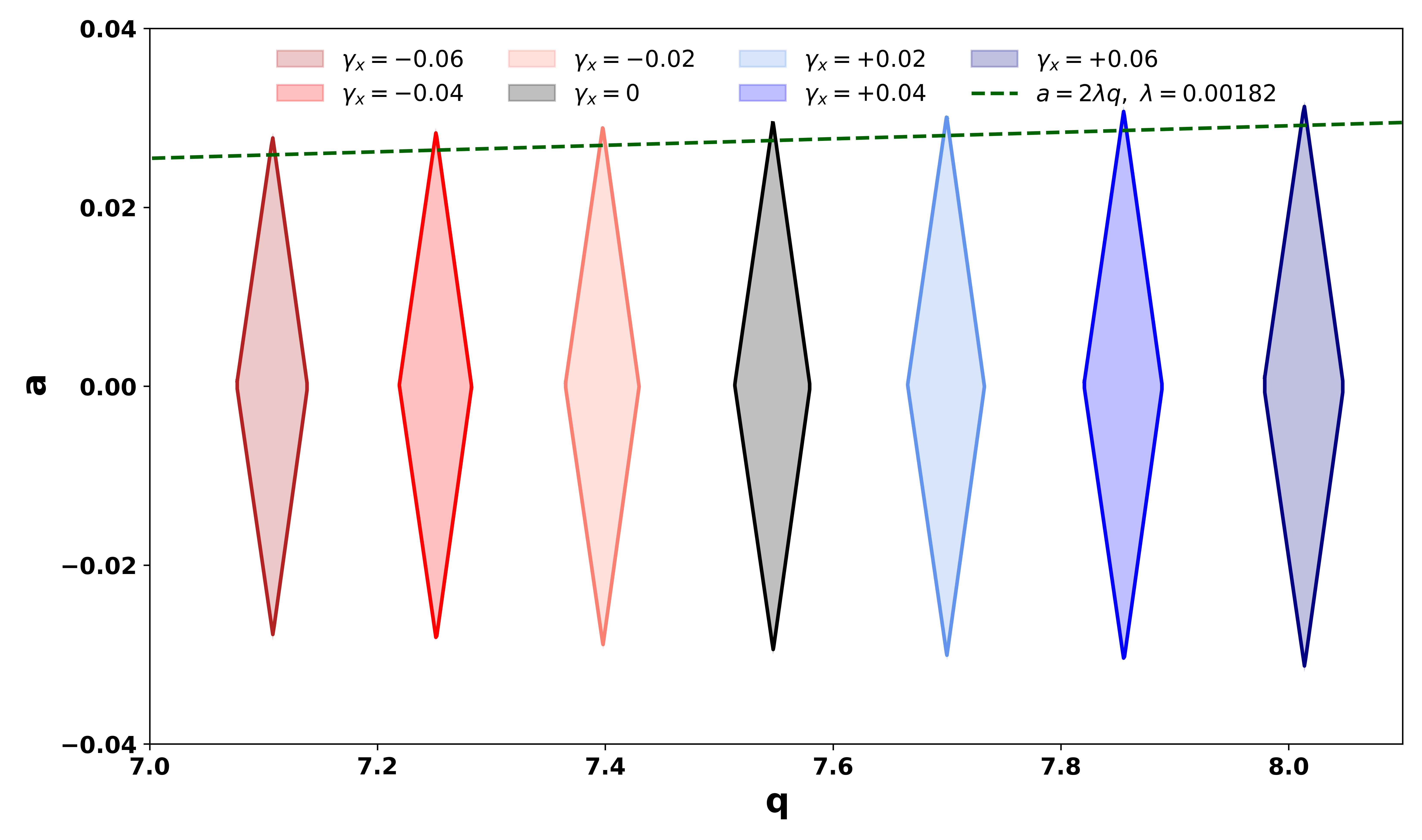}
    \caption{Second stability region of radially asymmetric QMFs for $\gamma_x=\pm0.06$, $\pm0.04$, $\pm0.02$ in comparison to a symmetric QMF ($\gamma_x=0$) as obtained from RK45 integration of the normalized equations of motion (eq.~\ref{eq5}) in the \((a,q)\) plane. Dashed line shows the scan parameter $\lambda = 0.00182$ with which the transmission performances were simulated.}
    \label{fig:2}
\end{figure}

Numerical integration of eq.~\ref{eq7} was carried out over the dimensionless interval $\xi=0$ to $200$ starting from the initial condition $(x,\dot x,y,\dot y) = (0,1,0,1)$. The stability of ion motion was determined by monitoring the evolution of all phase-space variables during integration. A trajectory was deemed stable if none of the coordinates or velocities exceeded a predefined bound of 200 at any point in time; otherwise, the motion was classified as unstable and the integration was terminated. 

The stability landscape was constructed by systematically scanning the parameter space in the vicinity of the second stability region. The Mathieu parameter $a$ was varied from $-0.04$ to $+0.04$, while $q$ was scanned from $7.00$ to $8.20$, with each parameter sampled using 1000 uniformly spaced values. All simulations were carried out for a fixed field radius $r_0 = 5$ mm and for several radial asymmetry parameters, $\gamma_x = \pm 0.02, \pm 0.04,$ and $\pm 0.06$.

The resulting second stability regions are shown in Figure.~2. For $\gamma_x > 0$, the stability zone is shifted toward higher $q$ values, whereas for $\gamma_x < 0$, it shifts toward lower $q$ values, consistent with the scaling relation given by eq.~8. The presence of radial asymmetry thus produces a systematic shift in the apex of the stability diagram, analogous to that previously reported for the first stability region~\cite{jana2025radial}.

\subsection{Octupole field}

The general form of the potential in a QMF with symmetrically displaced diagonal pair of electrodes is given by~\cite{jana2025radial} 

\begin{equation}\label{eq14}
\Phi(x,y,t) = \operatorname{Re}\!\sum_{N=0}^{\infty} A_{2N }\left(\frac{2z^2}{x_0^2+y_0^2}\right)^N (U-V_0\cos\omega t),
\end{equation}
where $z=x+iy$, and $A_{2N}$ denotes the dimensionless amplitude of the $2N$th-order multipole component.

Earlier simulation studies have reported multipole coefficients $A_{2N}$ for a range of rod-to-field radius ratios ($\eta$) and various asymmetry parameters~\cite{jana2025radial,JANA2025117495} where a monotonic increase in the octupole amplitude $|A_4|$ is demonstrated with increasing inward or outward displacement of a diagonal rod pair for $\eta = 1.10$, 1.12, and 1.14~\cite{jana2025radial}.

In the present work, transmission studies are carried out at $\eta = 1.13$, and the corresponding multipole field analysis is performed for the same $\eta$ over various radial asymmetry parameters using SIMION\cite{dahl2000simion} 3D simulations, following the methodology described in Ref.~\cite{jana2025radial}. The electrostatic potential surface inside the QMF is simulated using a grid unit (GU) of $0.1$~mm by applying a DC potential of $+1$~V to the electrode pair aligned along the $x$ direction and $-1$~V to the opposite pair aligned along the $y$ direction. The field radius is taken as $r_0 = 5$~mm. A cylindrical mu-metal shield of radius $4r_0$ is held at ground potential to define the simulation boundary. Potential simulations are performed by symmetrically displacing the $x$-pair electrodes corresponding to $\gamma_x = 0, \pm 0.02, \pm 0.04,$ and $\pm 0.06$, while keeping the rod radius fixed at $r=5.65$~mm. 

While the multipole coefficients $A_{2N}$ follow the same overall trends as reported earlier for $\eta=1.10, 1.12, 1.14$~\cite{jana2025radial}, specific values for $\eta=1.13$, $\gamma_x = 0$ and $\pm 0.04$ are listed in Table~1. Notably, the octupole coefficient $A_4$ is negative for inward displacement and positive for outward displacement of the $x$-pair rods, which are biased at a positive DC potential. An offset (monopole) component $A_0$ appears in the asymmetric configuration; it is positive for inward displacement and negative for outward displacement under the present voltage configuration. All $A_{2N}$ coefficients reverse sign upon interchanging the DC polarities applied to the $x$- and $y$-pair electrodes.

\begin{table*}[h!]
\centering \label{table2}
\caption{\label{asym-coeff}Amplitude $A_{2N}$ vs. asymmetric parameter($\gamma_x$) for an asymmetric setup of $\eta$=1.13}
\begin{tabular}{c|ccccccc}
\hline
$\gamma_x$ & $A_0 \times 10^2$ & $A_2$ & $A_4 \times 10^3$ & $A_6 \times 10^3$ & $A_8 \times 10^3$ & $A_{10} \times 10^3$ \\  
\hline
$-0.04$ & 4.2828 & 1.0429  &  -3.3323 &  -2.7810 & -3.3497 & -5.3255 \\
0 & 0  &  1.0016 &  0  &  0.9717 & 0 & -2.4594\\
+0.04 &  -4.1015 & 0.9609 & 0.9343 & 2.0887  & 0.7934 & -2.0069 \\
\hline
\end{tabular}
\label{table2}
\end{table*}

\section{QMF performance} 

The performance of a quadrupole mass filter operating in the second stability zone with an added octupole component is analyzed through ion transmission simulations using SIMION by systematically displacing the $x$-pair electrodes inward or outward. The performance of a QMF, even in a symmetric configuration, is governed by several parameters, particularly the rod-to-field radius ratio ($\eta$) and the scan parameter $\lambda = a/2q$. Prior to introducing radial asymmetry, a series of simulations was performed to identify suitable values of $\eta$ and $\lambda$.

All simulations presented in this work were carried out using a monoenergetic ion beam consisting of 2000 ions (in some cases increased to 5000, or even 10000 when transmission was low), each with a mass-to-charge ratio of $40$~u/C and a longitudinal kinetic energy of $2$~eV. The ions were assigned a time-of-birth uniformly distributed between 0 and $2~\mu$s and were initially distributed uniformly over a circular area of radius $0.02r_0$ at the entrance of a QMF of length $160$~mm, operated at a drive frequency of $500$~kHz.

\subsection{Transmission characteristics}
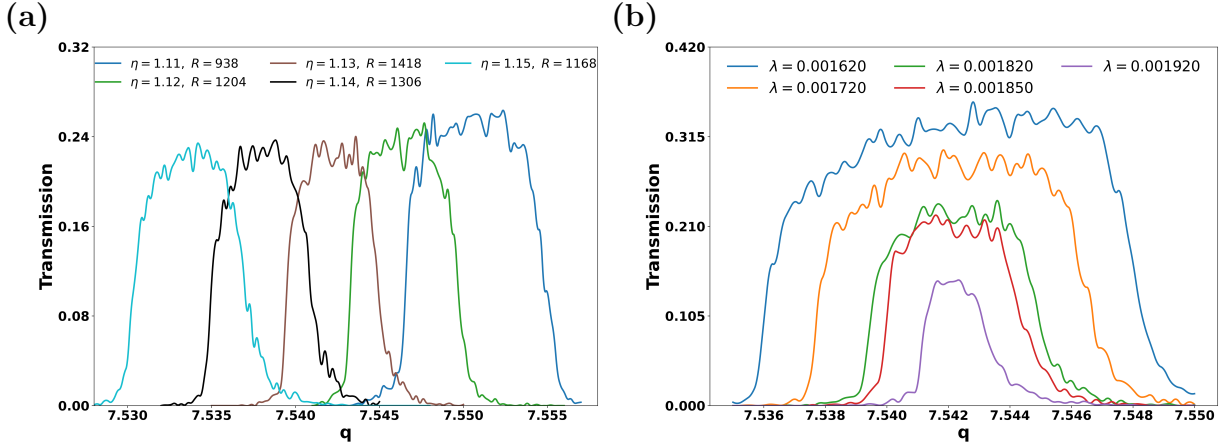
\begin{figure}[H]
    \centering
    \begin{subfigure}[b]{0.48\textwidth}
        \begin{picture}(0,0)
            \put(-8,160){\textbf{(a)}}
        \end{picture}
        \includegraphics[width=\textwidth]{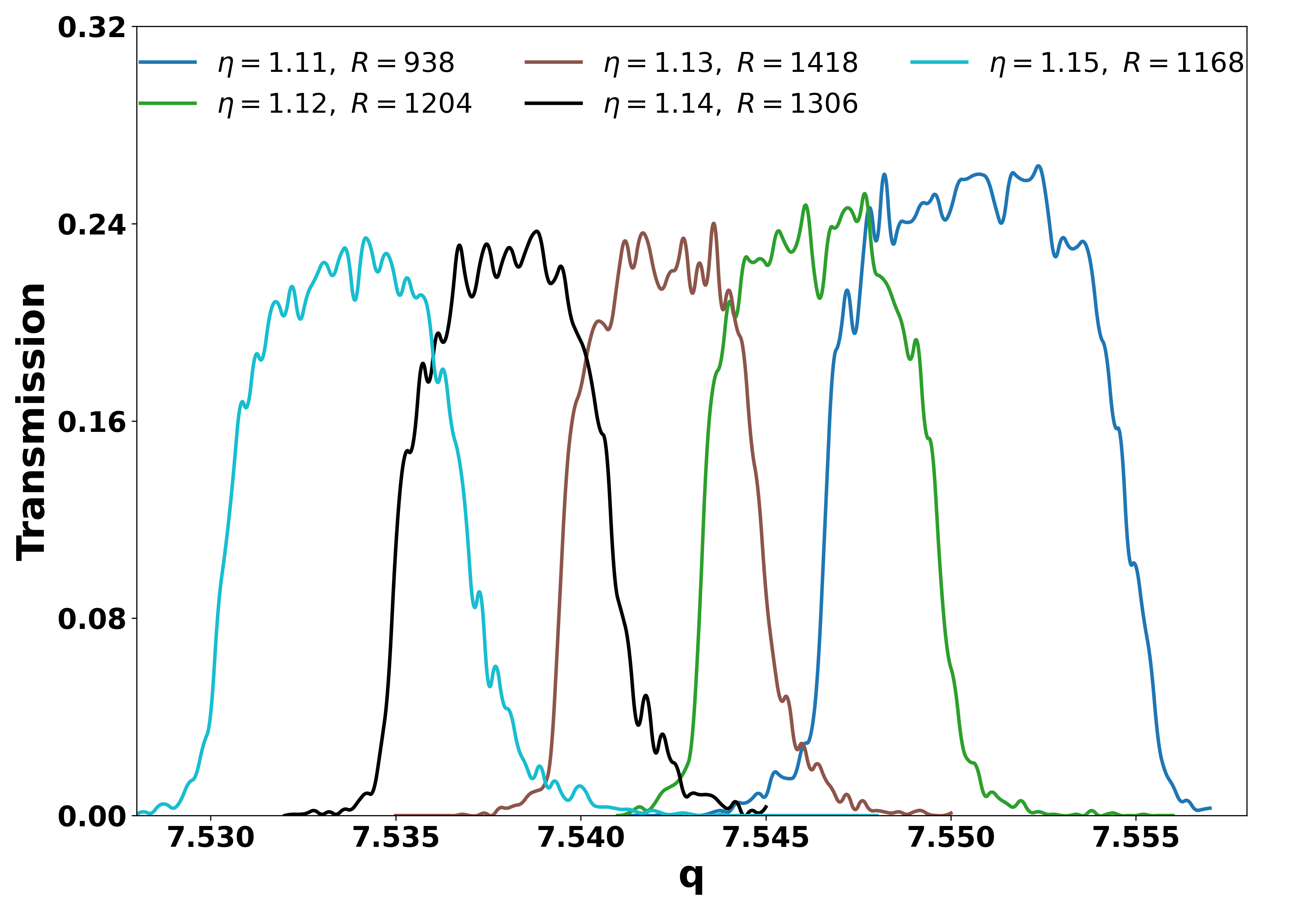}
        \label{fig:a}
    \end{subfigure}
    \begin{subfigure}[b]{0.48\textwidth}
        \begin{picture}(0,0)
            \put(-8,160){\textbf{(b)}}
        \end{picture}
        \includegraphics[width=\textwidth]{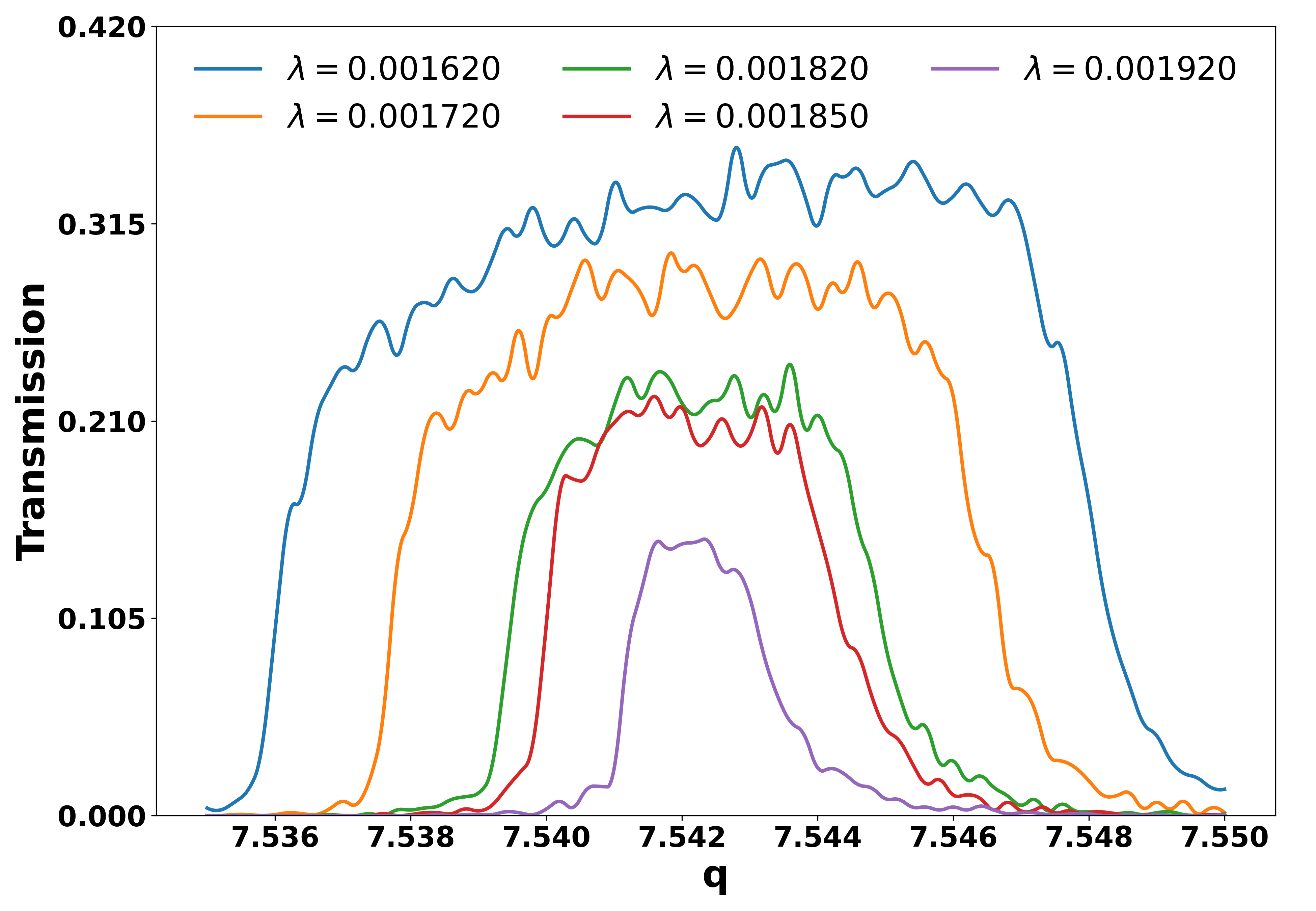}
        \label{fig:b}
    \end{subfigure}
    \caption{Transmission characteristics of symmetric QMF for (a) different $\eta$ at a fixed $\lambda = 0.00182$, and (b) different scan parameter ($\lambda$) at a fixed $\eta=1.13$ in the second stability zone.}
    \label{fig:3}
\end{figure}

Figure~\ref{fig:3}(a) shows the transmission contours corresponding to $\eta = 1.11$, 1.12, 1.13, 1.14 and 1.15 at a fixed scan parameter $\lambda = 0.00182$. Noticeable differences are observed in the transmission characteristics for different $\eta$ values, which are commonly attributed to the interplay between the dodecapole and icosapole components of the radial confinement potential~\cite{hogan2009effects,schulte1999nonlinear}. The transmission peak exhibits a systematic shift, accompanied by variations in the contour width. Among the cases considered, the transmission contour at $\eta = 1.13$ appears comparatively sharper, indicating improved mass resolution. Consequently, all subsequent studies are carried out at $\eta = 1.13$.

Figure~\ref{fig:3}(b) presents the transmission contours for different values of the scan parameter $\lambda$ at a fixed rod-to-field radius ratio of $\eta = 1.13$. As expected, increasing $\lambda$ results in progressively narrower contours, corresponding to improved mass resolution, albeit at the expense of reduced transmission efficiency. A reasonable compromise between resolution and transmission is obtained at $\lambda = 0.00182$, where the peak transmission exceeds $20\%$ while maintaining a relatively sharp profile. Accordingly, all subsequent studies are carried out at $\lambda = 0.00182$.
\vspace{2mm}
 
\begin{figure}
    \centering

    \begin{minipage}{1.0\textwidth}
        \centering
        \includegraphics[width=\linewidth]{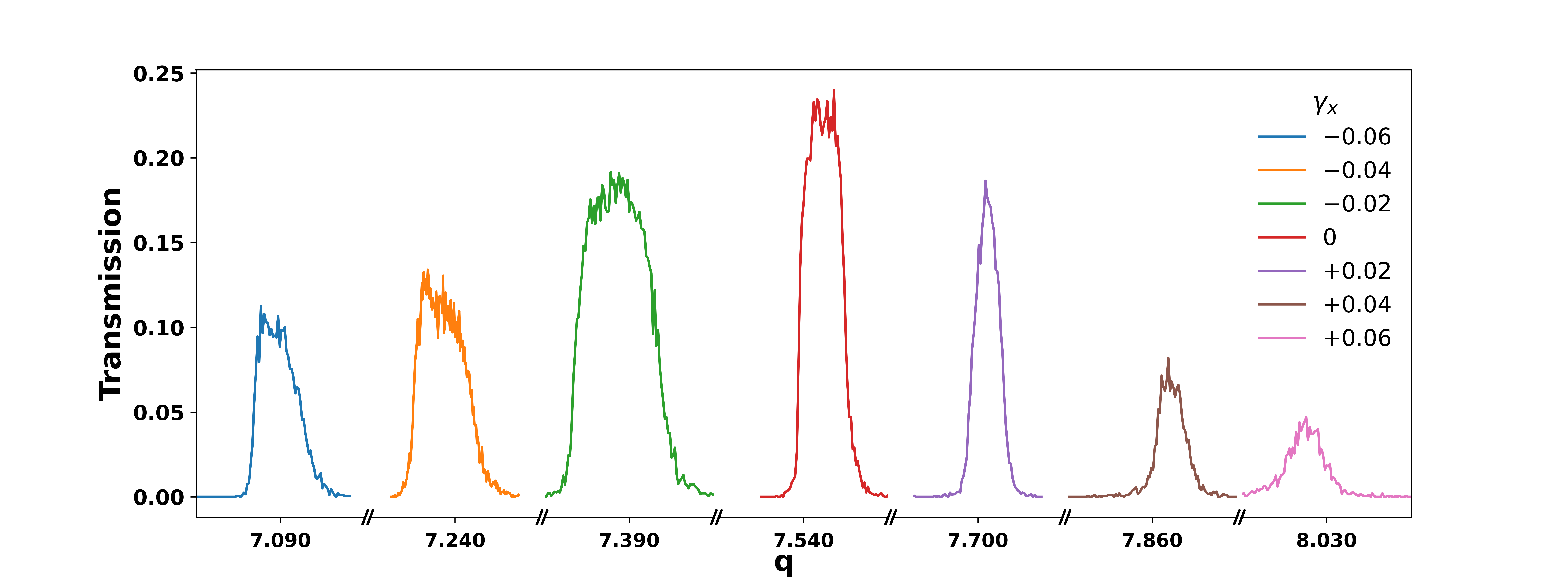}
        \caption*{\small }
    \end{minipage}
    
    \caption{ Transmission profile for symmetric ($\gamma_x= 0$) and asymmetric QMFs ($\gamma_x = \pm0.02, \pm0.04, \pm0.06$) at $\eta=1.13$ and $\lambda=0.00182$ with $+U$ applied to the displaced electrodes.}
    \label{fig:4}
\end{figure}

Radial asymmetry in the transverse plane is introduced by symmetrically displacing the $x$-pair electrodes from their normal positions by an amount $\Delta x = \gamma_x r_0$ and the transmission characteristics are simulated for $\gamma_x = \pm 0.02, \pm 0.04,$ and $\pm 0.06$ under the voltage configuration shown in Figure.~\ref{fig:1}(b), with the $x$-pair of rods biased at a positive DC potential ($+U$) and the $y$-pair biased at a negative DC potential ($-U$). As shown in Figure.~\ref{fig:4}, the transmission peak shifts monotonically with the asymmetry parameter, consistent with the displacement of the stability apex caused by the modified quadrupole field in the asymmetric configuration (Figure.~\ref{fig:2}). The transmission profiles exhibit changes in shape, accompanied by a pronounced reduction in transmission efficiency with increasing asymmetry.

The transmission characteristics of a QMF operating in the first stability zone have been reported to be sensitive to the polarity of the applied DC potential in an asymmetric configuration~\cite{ding2003quadrupole}. To examine the effect of DC polarity on transmission in the second stability zone, the present study is repeated by interchanging the polarity of the DC potential applied to the $x$ and $y$-pair electrodes. While the shift in the transmission peak and the reduction in transmission efficiency with increasing asymmetry follow the same overall trends as observed earlier, the shape of the transmission contours exhibits noticeable differences compared to the earlier polarity configuration.

\begin{figure}[H]
    \centering
    \begin{subfigure}[b]{0.48\textwidth}
        \begin{picture}(0,0)
            \put(-8,160){\textbf{(a)}}
        \end{picture}
        \includegraphics[width=\textwidth]{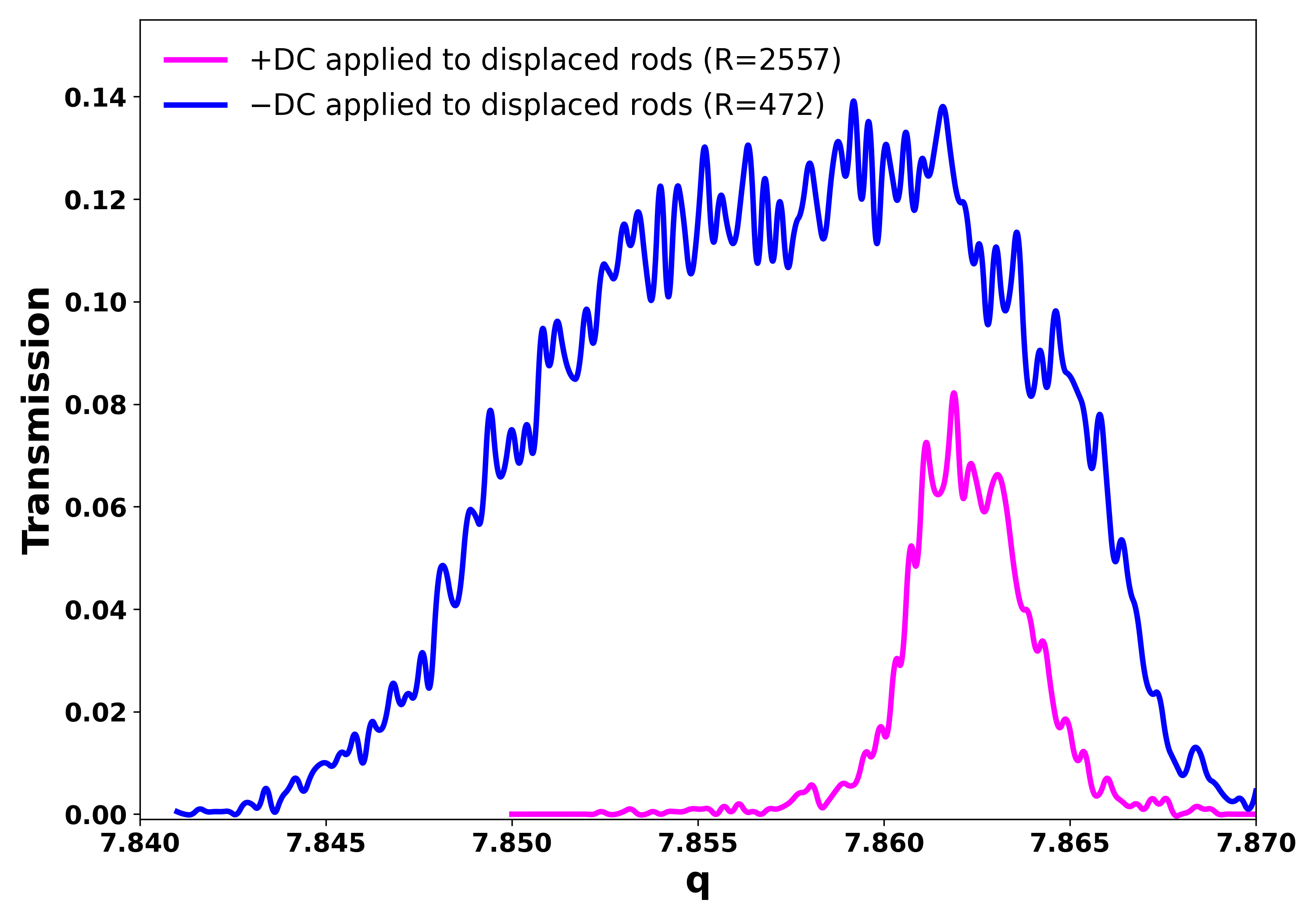}
       \label{fig:a}
    \end{subfigure}
    \hspace{2mm} 
    \begin{subfigure}[b]{0.48\textwidth}
        \begin{picture}(0,0)
            \put(-8,160){\textbf{(b)}}
        \end{picture}
        \includegraphics[width=\textwidth]{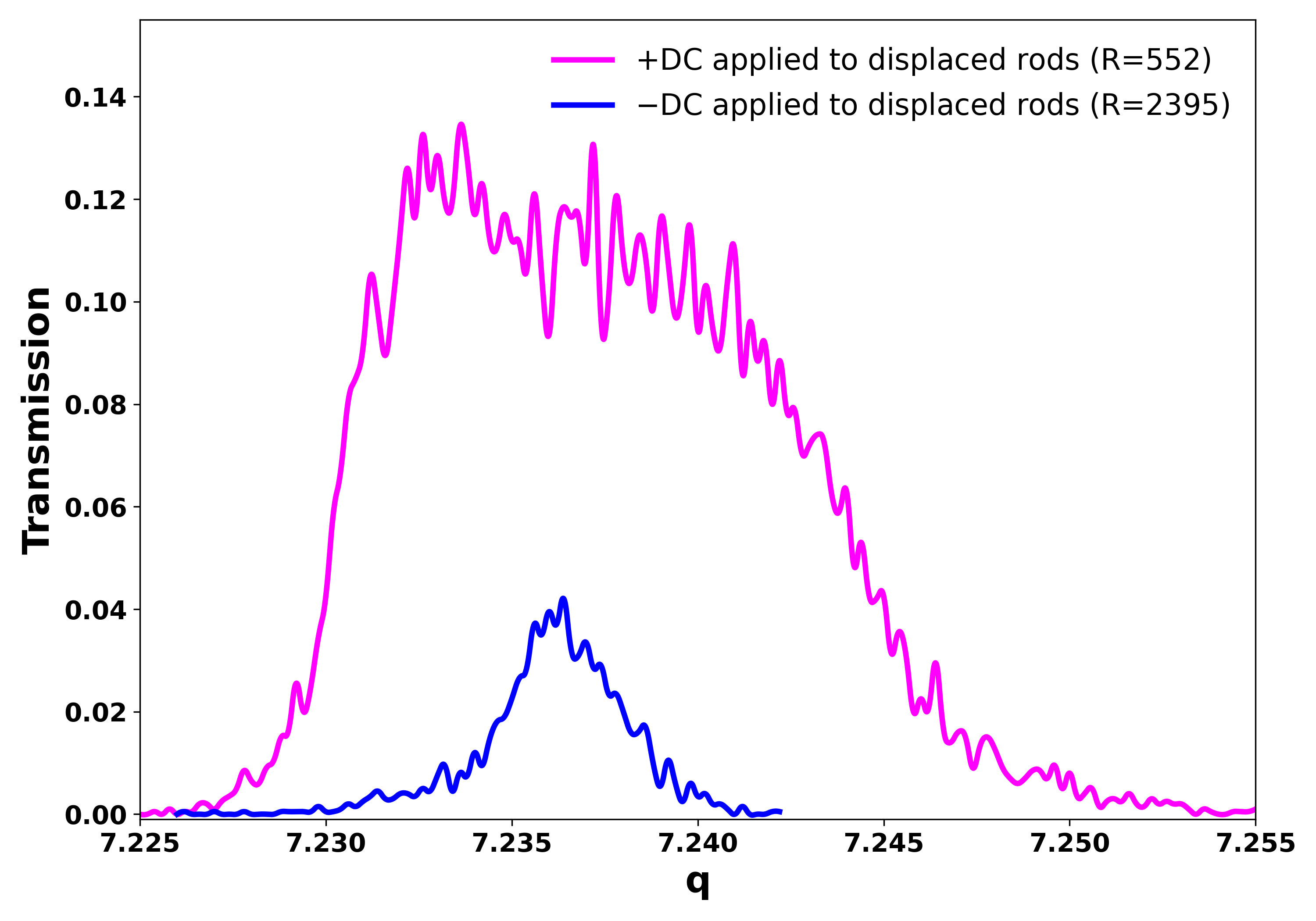}
      \label{fig:b}
    \end{subfigure}
    \caption{Transmission profile for $\pm U$ applied to the displaced rod pair in the asymmetric configuration of (a) $\gamma_x=+0.04$ and (b) $\gamma_x=-0.04$.}
    \label{fig:5}
\end{figure}

Figures~\ref{fig:5}(a) and (b) present a direct comparison of the transmission characteristics corresponding to $+U$ and $-U$ set at the displaced rods of radial asymmetry $\gamma_x = +0.04$ and $=-0.04$ respectively. As shown in Figure.~\ref{fig:5}(a), outward displacement of the $x$-pair electrodes ($\gamma_x = +0.04$) biased at $+U$ results in a narrower transmission contour accompanied by reduced transmission efficiency compared to $-U$ set at these electrodes. In contrast, for inward displacement of the $x$-pair electrodes ($\gamma_x = -0.04$), a similarly narrower contour with reduced transmission is observed when the displaced rods are biased at $-U$, as shown in Figure.~\ref{fig:5}(b). 

\subsection{Resolution}

In order to draw a quantitative comparison of the asymmetric QMF for various asymmetry factor and polarity configurations applied to the displaced rods, we define a resolution parameter $R=q/\Delta q$ in reference to the transmission contours, where $q$ corresponds to the transmission peak and $\Delta q$ is the associated FWHM.

\begin{figure}[H]
    \centering
    \includegraphics[width=1.0\linewidth]{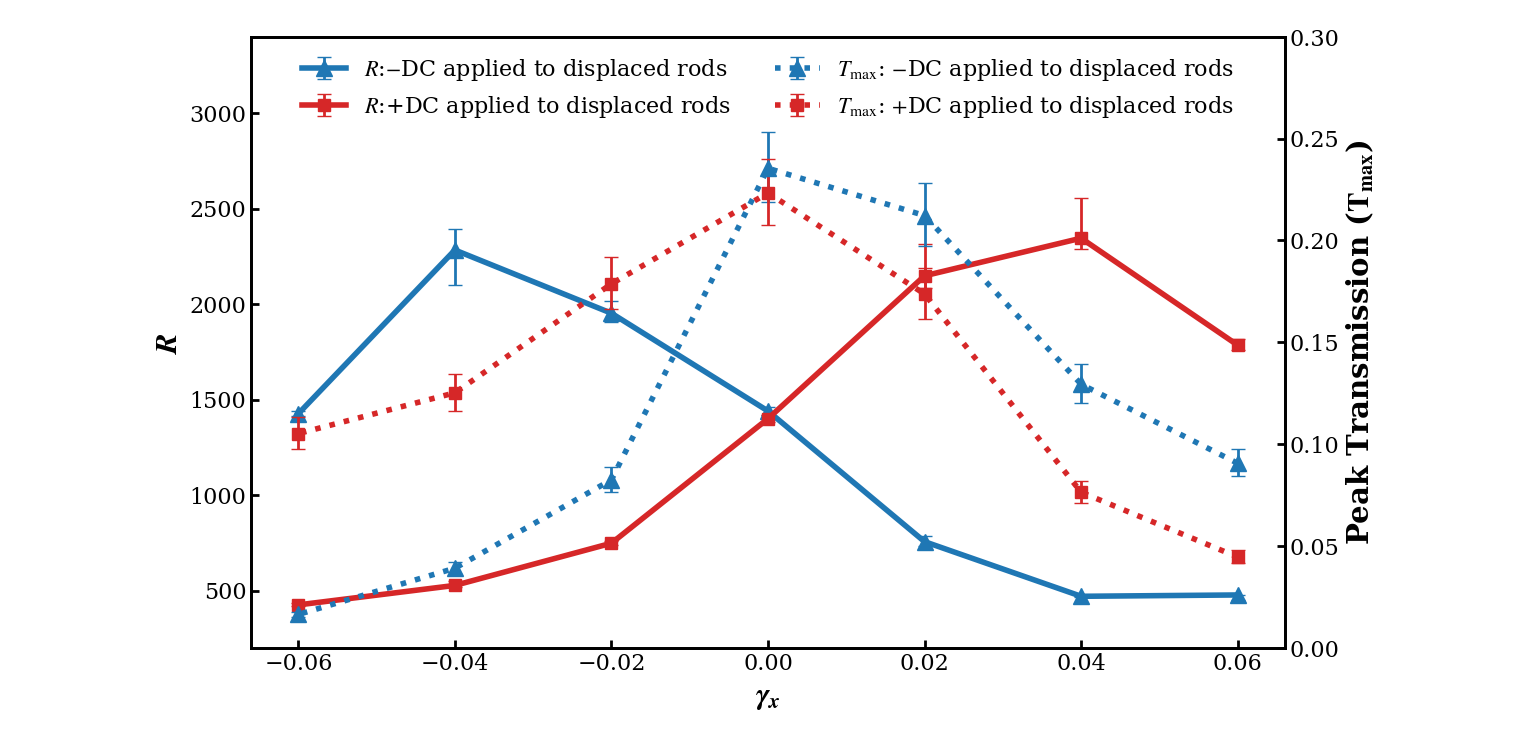}
    \caption{Resolution $R$ (solid lines) and peak transmission (dotted lines) as a function of the radial asymmetry parameter $\gamma_x=0, \pm0.02, \pm0.04, \pm0.06$. $R$ is determined using a 15\% transmission window around the peak, where the middle transmission level serves as the reference while the upper and lower bounds define the uncertainty in $T_{max}$. The full width at half maximum is evaluated at 50\% of each transmission level, and the resulting spread in resolution determines the error bar in $R$.}
    \label{fig:6}
\end{figure}

Figure~\ref{fig:6} shows the systematic variation of transmission resolution and efficiency as a function of the radial asymmetry factor for $\eta = 1.13$, for both polarities of the DC potential applied to the displaced electrodes. When the displaced $x$-pair electrodes are biased at $+U$, the resolution decreases monotonically for inward displacements and increases for outward displacements, attaining a maximum at $\gamma_x = +0.04$. Upon reversal of the DC polarity, a complementary behavior is observed: the resolution decreases monotonically with increasing outward displacement and increases for inward displacement, reaching a maximum at $\gamma_x = -0.04$.

\begin{figure}[H]
    \centering
    \begin{subfigure}[b]{0.48\textwidth}
        \begin{picture}(0,0)
            \put(2,163){\textbf{(a)}}
        \end{picture}
        \includegraphics[width=\textwidth]{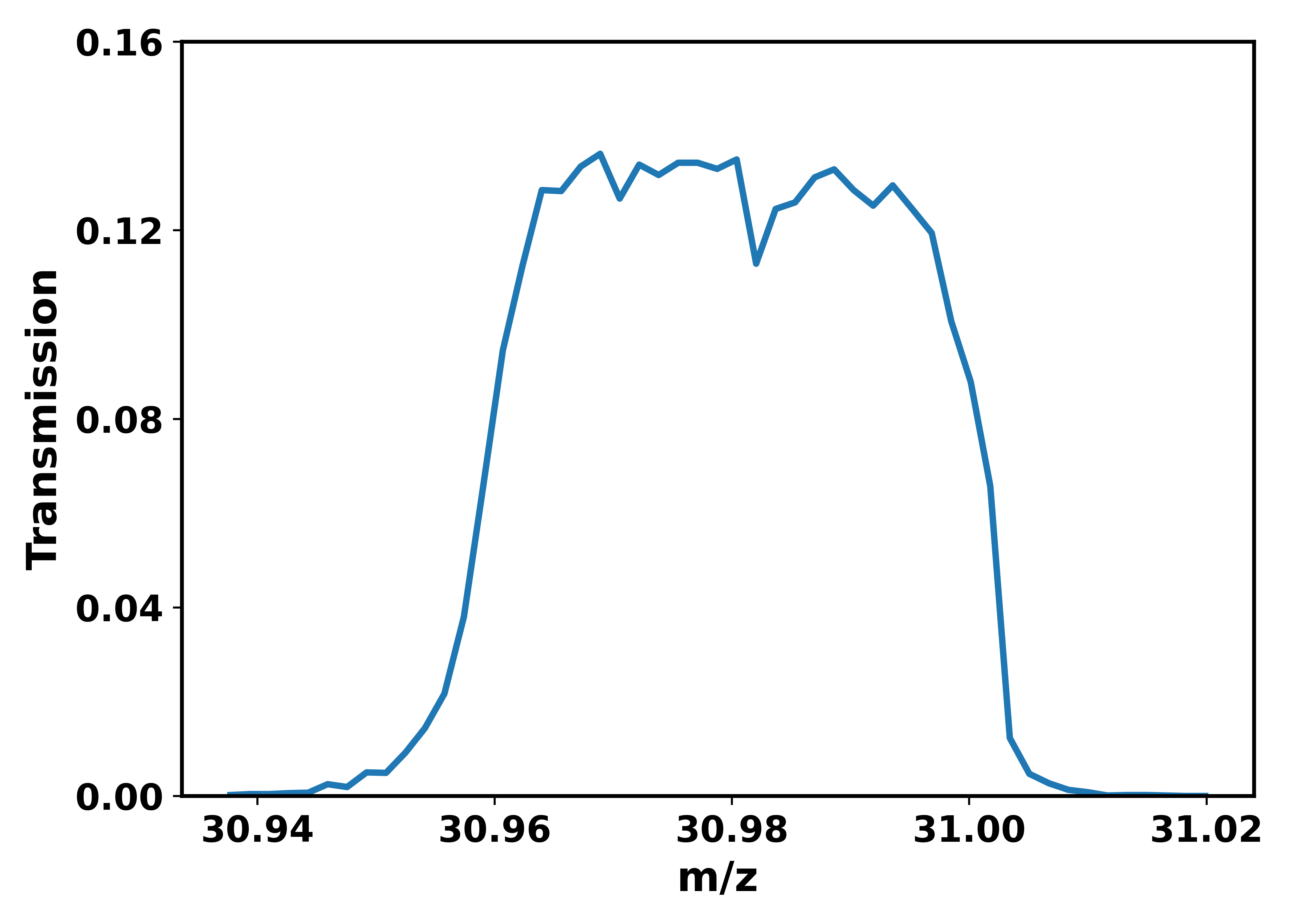}
     \label{fig:a}
    \end{subfigure}
    \hspace{1mm} 
    \begin{subfigure}[b]{0.48\textwidth}
        \begin{picture}(0,0)
            \put(2,163){\textbf{(b)}}
        \end{picture}
        \includegraphics[width=\textwidth]{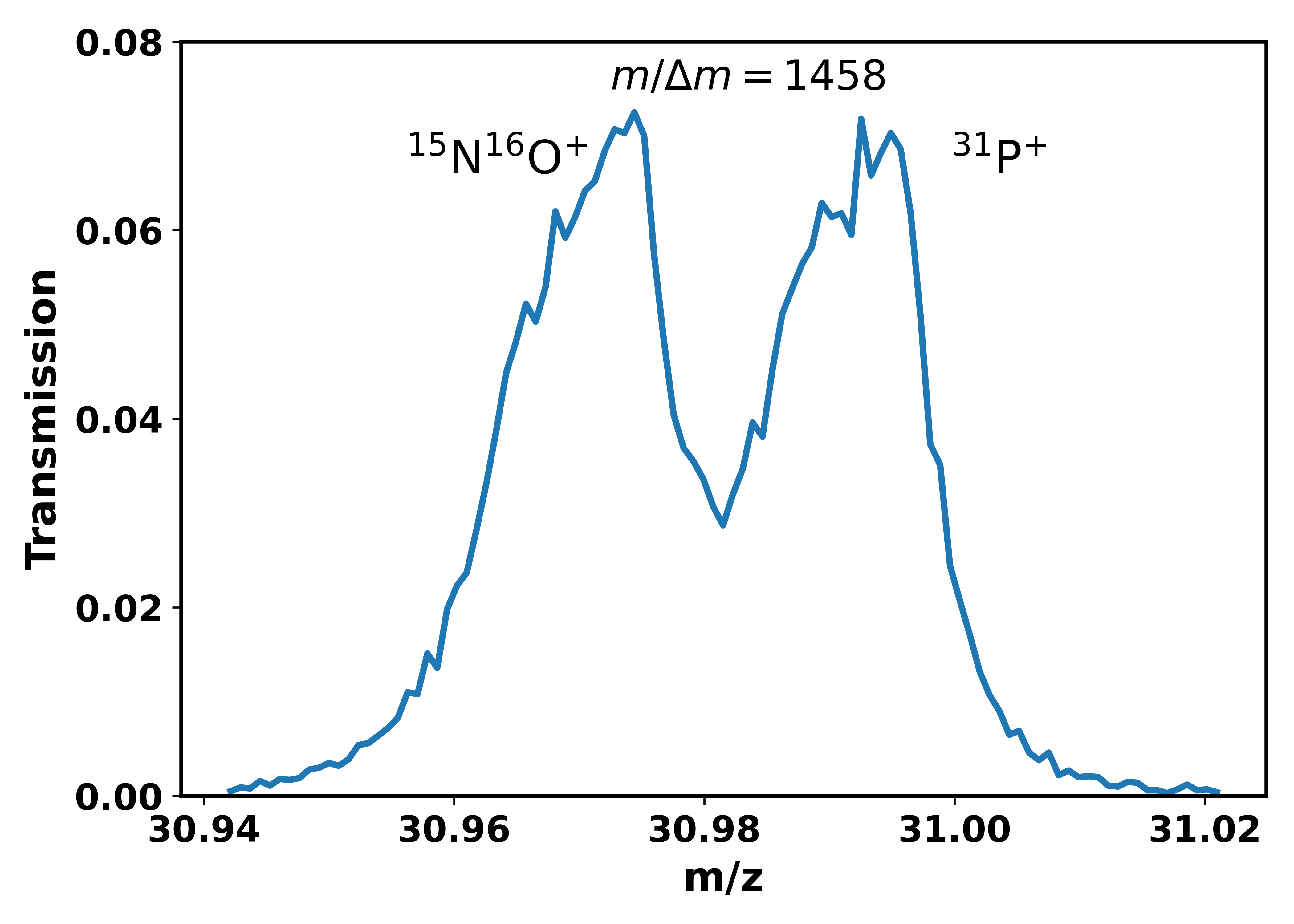}
      \label{fig:b}
    \end{subfigure}
    \caption{Transmission of $^{15}\mathrm{N}^{16}\mathrm{O}^{+}$
 and $^{31}\mathrm{P}^{+}$ through (a) symmetric QMF where the ion species are not resolved, and (b) asymmetric QMF ($\gamma_x=+0.04$, with displaced rod pair at $+U$) showing the ion species resolved. The species are also found to be resolved in $\gamma_x=-0.04$, with displaced rod pair at $-U$.}
    \label{fig:7}
\end{figure}

Motivated by the enhanced resolution observed at $\gamma_x = 0.04$ with the $x$-rod pair biased at $+U$, the mass resolving performance of the QMF was further examined using two near-degenerate ion species, $^{15}\mathrm{N}^{16}\mathrm{O}^{+}$ and $^{31}\mathrm{P}^{+}$. As shown in Figure.~\ref{fig:7}(a), the two species remain unresolved in the symmetric configuration, whereas Figure.~~\ref{fig:7}(b) exhibits a clear separation between them. This result confirms that the enhanced resolving capability observed in Figure.~~\ref{fig:6} is a direct consequence of engineered radial asymmetry ($\gamma_x = 0.04$) and not an artifact of the simulation.

\section{Discussions}

A key observation from Figure.~\ref{fig:6} is the complementary behavior of the $R$–$\gamma_x$ characteristic under reversal of the DC polarity. For a given asymmetry parameter, the mass resolution exhibits a pronounced dependence on the polarity applied to the displaced rod pair. For example, at $\gamma_x = 0.04$, reversing the DC polarity leads to an enhancement in resolution by approximately a factor of six.

A symmetric quadrupole configuration inherently contains higher-order multipole components, most prominently a dodecapole term ($A_{6}$) and a smaller icosapole contribution ($A_{10}$). The transmission characteristics are found to be invariant under reversal of the DC polarity applied to the electrodes. This invariance indicates that, the sign of $A_{6}$ (and $A_{10}$) does not play a significant role in determining either the stability boundary or the transmission performance of symmetric QMF in the second stability zone.

The situation changes markedly when radial asymmetry is introduced by displacing a pair of electrodes. In this case, additional multipole components, predominantly the octupole term ($A_4$), along with a smaller hexadecapole contribution $A_8$ appear in the field expansion. Reversal of the DC polarity leads to a reversal in the sign of $A_4$, resulting in a polarity-dependent shift of the stability boundary unlike the symmetric case.

\subsection{Stability boundary shift}
While Ding et al. demonstrated that an added octupole field can either contract or expand the first stability zone depending on the applied DC polarity, thereby enabling enhanced mass resolution~\cite{ding2003quadrupole}, a similar mechanism is expected to operate in the second stability zone. However, in radially asymmetric QMFs, the second stability region undergoes a substantial displacement due to modification of the underlying quadrupole potential itself (Figure.~\ref{fig:2}), which complicates a direct assessment of the isolated contribution of the octupole component relative to the symmetric configuration. In the present work, the effect of DC polarity reversal—equivalently, the role of the sign of the octupole coefficient—is therefore examined in detail through SIMION simulations of the stability boundary in the second stability zone for asymmetric QMF configurations.

\begin{figure}
    \centering

    \begin{subfigure}{\linewidth}
        \centering
        \includegraphics[width=0.8\linewidth]{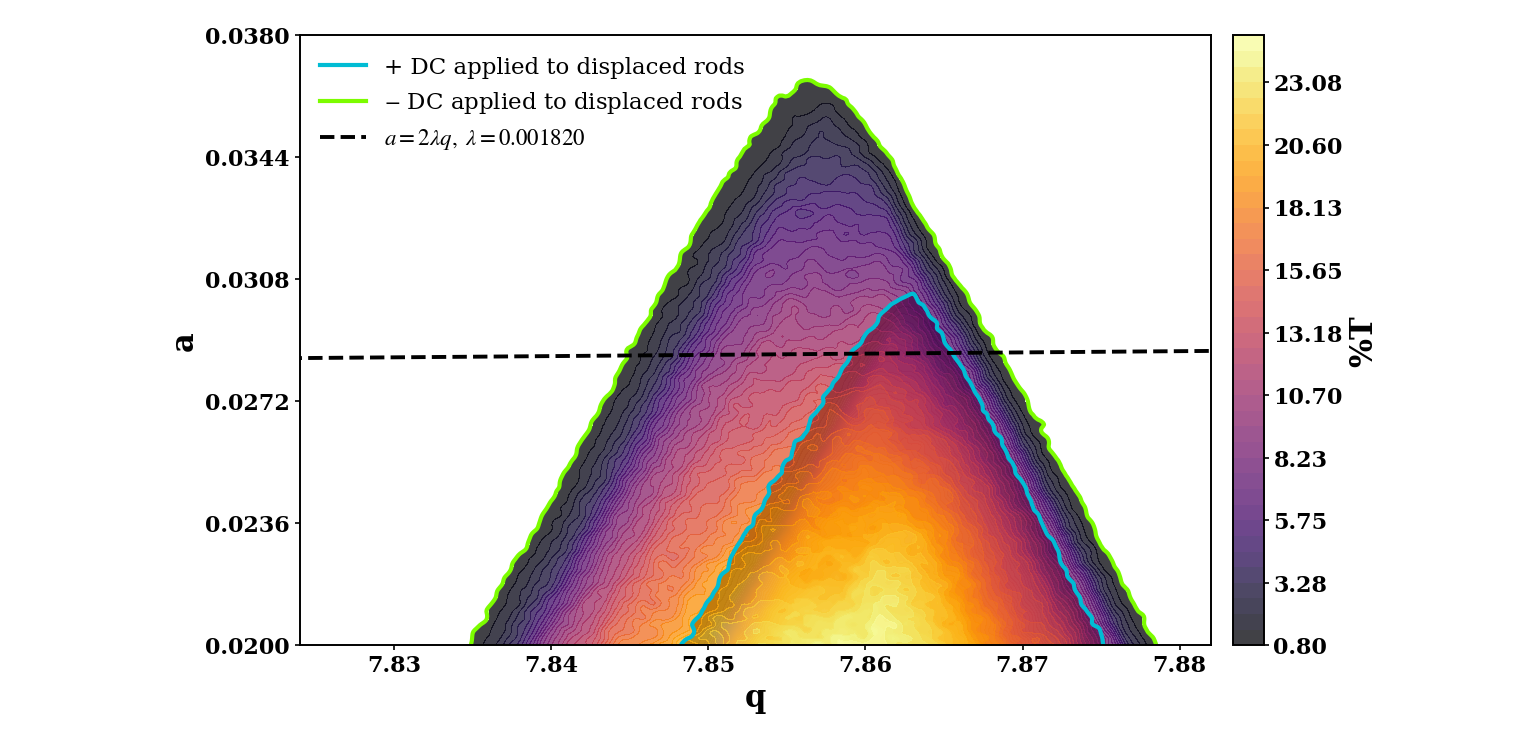}
        \label{fig:stability_posneg}
    \end{subfigure}
    \caption{Tip of the second stability zone of a radially asymmetric QMF ($\gamma_x = +0.04$) with displaced electrode pair set at $\pm U$. The stability boundary is defined at 0.8\% transmission. Dashed line corresponds to $\lambda = 0.00182$.}  
    \label{fig:8}
\end{figure}

To trace the second stability zone of the asymmetric QMF, simulations were performed using $5000$ ions under the same operating conditions employed for the transmission studies described earlier. The Mathieu parameters $a$ and $q$ were varied in steps of $0.0002$ and $0.0004$, respectively. For each $(a,q)$ pair, an identical ensemble of ions with the same initial phase-space configuration was propagated through the QMF, and the corresponding transmission percentage was evaluated. Figure~\ref{fig:8} presents the resulting stability heat maps in the $a$--$q$ plane, focusing on the region relevant to operation in the second stability zone for an asymmetric QMF with $\gamma_x = 0.04$. The stability boundary is defined by a transmission threshold of $0.8$\%.

As evident from Figure.~\ref{fig:8}, the stability region is wider when the displaced rod pair is biased at a negative DC voltage compared to positive DC biasing. Consequently, for a fixed scan line, the baseline resolution is higher and transmission efficiency is lower when a positive DC bias is applied to the displaced electrode pair. A similar behavior is observed for the asymmetric configuration with $\gamma_x = -0.04$, where the stability region is wider when the displaced rod pair is biased at a positive DC voltage, indicating enhanced resolution with reduced transmission efficiency for negative DC biasing of the displaced rod pair.
   
\subsection{RF phase selectivity}

\begin{figure}
    \centering
    \includegraphics[width=0.6\linewidth]{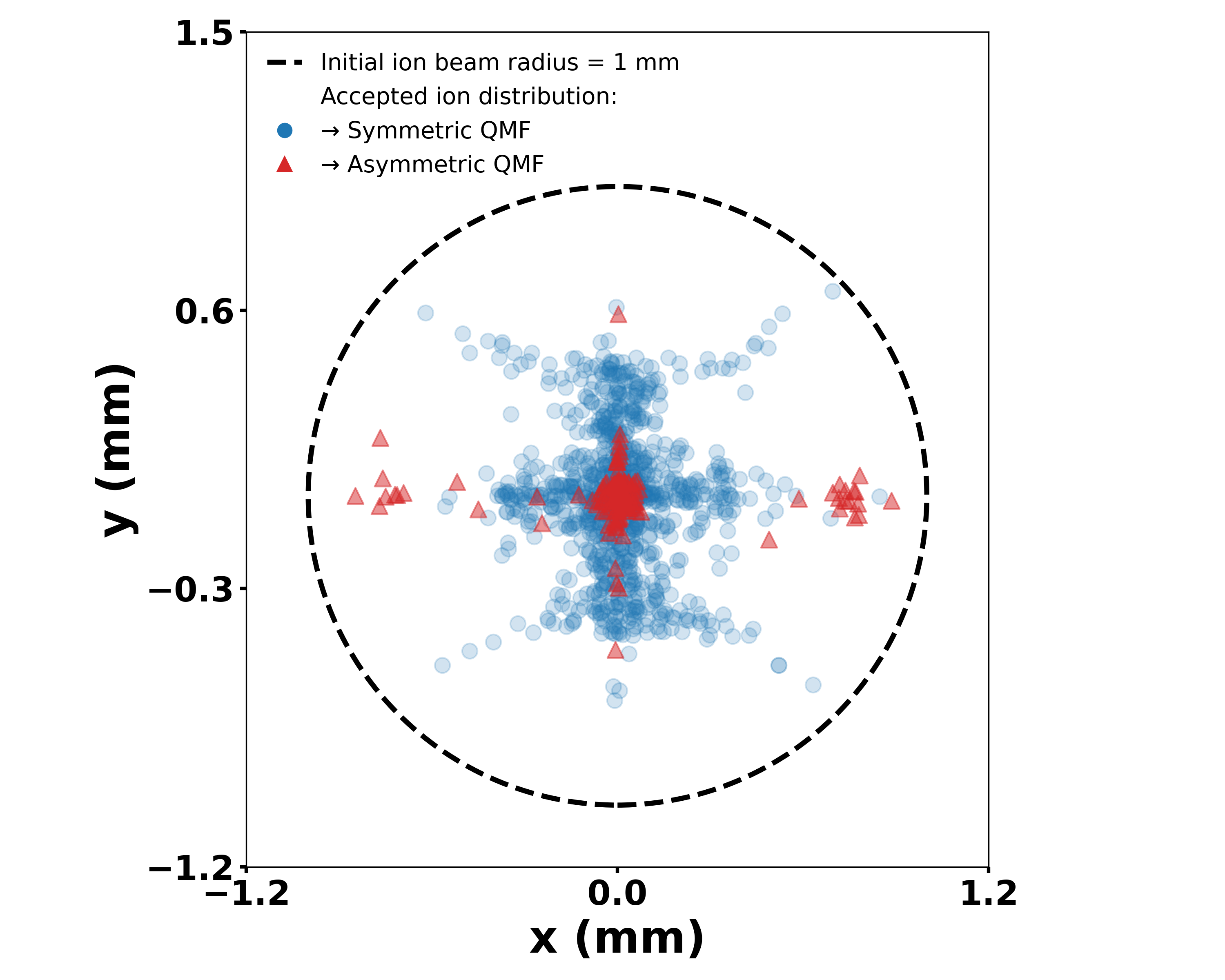}
    \caption{Scattered points showing the initial distribution of ions that pass through the symmetric and asymmetric QMF operated at peak transmission condition.}
    \label{fig:9}
\end{figure}

The enhanced resolution observed in the asymmetric configuration ($\gamma_x = 0.04$) with the displaced electrodes biased at a positive DC potential is accompanied by a significant reduction in transmission efficiency. The presence of the octupole component in the asymmetric mass filter reduces the effective acceptance of the QMF. Simulations were performed to compare the maximum spatial acceptance of a parallel ion beam for the symmetric and asymmetric QMFs. An ensemble of $10^{5}$ ions, initially distributed uniformly over a circular area of radius $0.2r_0$ with a time-of-birth of $0$–$2~\mu$s, was used in both cases. The operating parameters $(a,q)$ were chosen to correspond to the respective peak transmission conditions. The resulting initial ion distributions that lead to stable transmission are shown in Figure.~\ref{fig:9} for the symmetric and asymmetric ($\gamma_x = 0.04$, $x$-rod pair biased at $+U$) configurations, respectively. The reduced spatial acceptance observed for the asymmetric QMF persists across other operating points, indicating that radial asymmetry diminishes the QMF acceptance and thereby accounts for the observed reduction in transmission efficiency.

The reduced beam acceptance observed in the asymmetric QMF can be understood in terms of RF phase selectivity at the entrance of the mass filter. As shown by Dawson~\cite{dawson1975acceptance}, the acceptance leading to stable ion trajectories oscillates at the RF frequency, rendering the transmission probability highly sensitive to the RF phase angle ($\theta$) at the moment of ion injection. As further established by Douglas and Konenkov~\cite{konenkov20174d,turner2005effect}, this phase dependence becomes increasingly restrictive at higher resolution, where the acceptance ellipses contract, thereby reducing the fraction of ions transmitted through the mass filter.

To explicitly examine the role of RF phase at ion entry, the simulations were extended using the same number of ions ($10^{5}$) and identical operating conditions, except that all ions were injected simultaneously (time-of-birth $=0$). An offset phase ($\theta$) was introduced in the driving RF and systematically varied from $0$ to $2\pi$. Figures~10(a) and 10(b) show the initial spatial extent of the ion beam along the $x$ and $y$ directions that leads to stable transmission for the symmetric and asymmetric ($\gamma_x = 0.04$, $x$-rod pair biased at $+U$) configurations, respectively. In both cases, ions with a broader initial spatial distribution along either transverse direction are transmitted only over a narrow range of RF phase. The asymmetric QMF exhibits an even narrower admissible phase window, indicating a reduced acceptance associated with the enhanced resolution.

\begin{figure}
    \centering
    \begin{subfigure}{0.35\textwidth}
        \begin{picture}(0,0)
            \put(-15,116){\textbf{(a)}}
        \end{picture}
        \includegraphics[width=\textwidth]{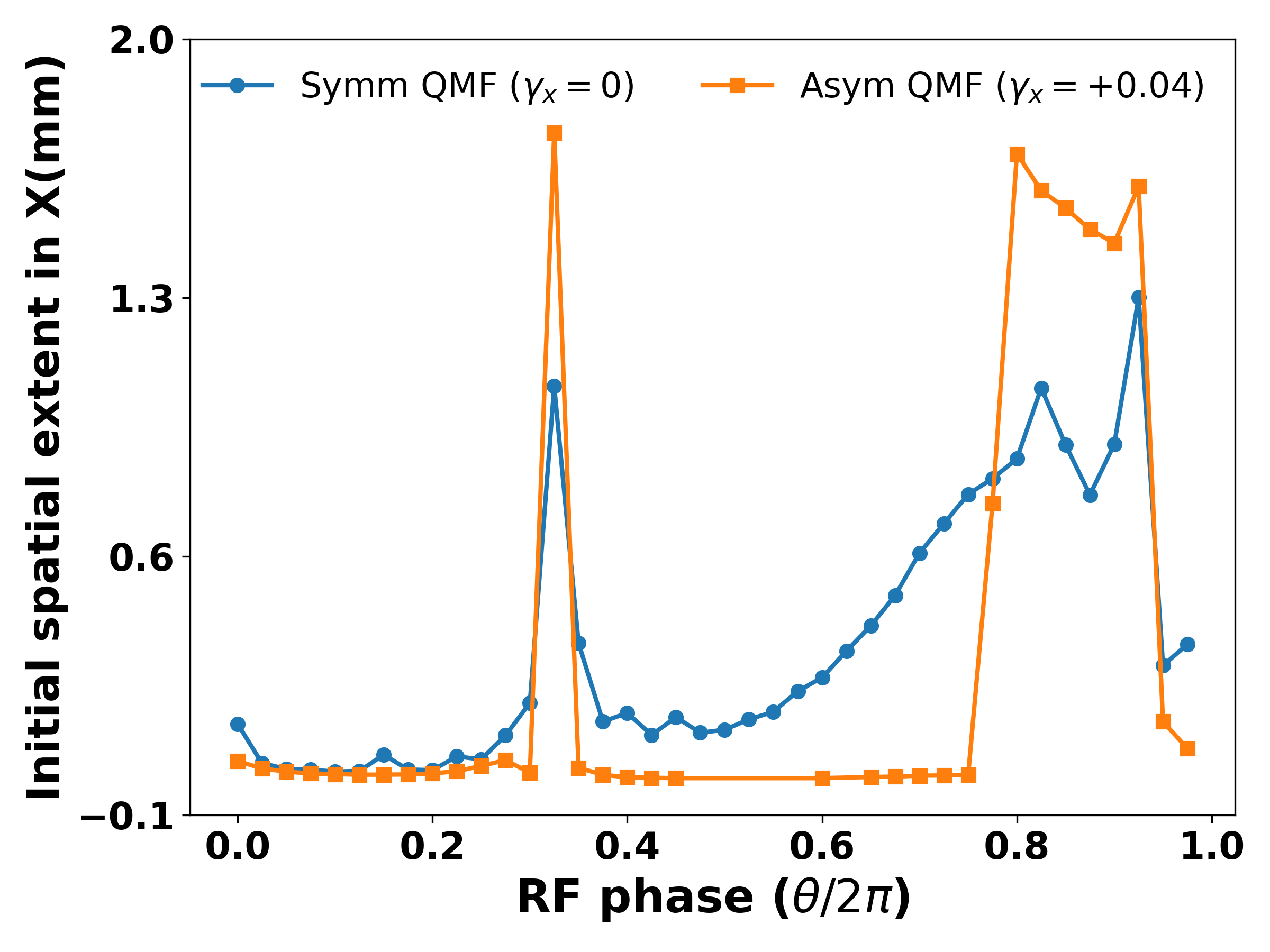}
        \label{fig:10a}
    \end{subfigure}

    \begin{subfigure}{0.35\textwidth}
        \begin{picture}(0,0)
            \put(-15,116){\textbf{(b)}}
        \end{picture}
        \includegraphics[width=\textwidth]{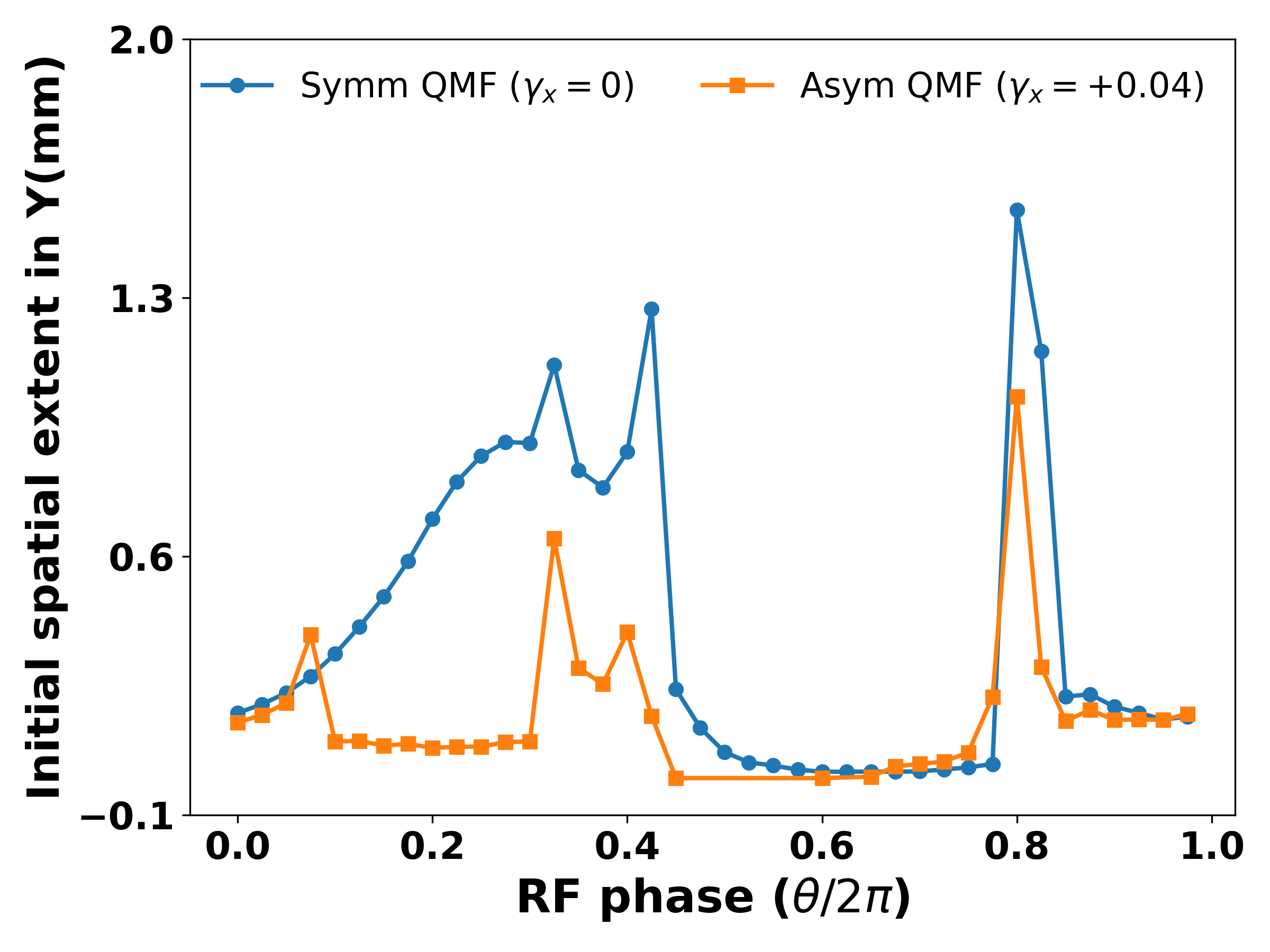}
        \label{fig:10b}
    \end{subfigure}

    \caption{Initial spatial extent along (a) $X$ and (b) $Y$ as a function of the initial RF phase for the symmetric and asymmetric QMFs.}
    \label{fig:10}
\end{figure}

An important observation in the asymmetric configuration is the pronounced broadening of the ion time-of-flight (TOF) through the mass filter. For ions with $m/e = 40$~u/C and a longitudinal energy of $2$~eV, the TOF through the symmetric QMF of the same length is approximately $53~\mu$s. In contrast, for the asymmetric setup, the TOF spans a wider range, extending from about $28~\mu$s to $60~\mu$s. The broadening in TOF may be attributed to octupole component and the enhanced RF phase selectivity.

\section{Conclusion}
In this work, the influence of the octupole field introduced by the symmetric displacement of a diagonally opposite electrode pair in a quadrupole mass filter has been systematically investigated, with particular emphasis on operation in the second stability zone. In addition to generating an octupole component, radial asymmetry modifies the effective quadrupolar potential, resulting in a shift of the stability apex and the corresponding transmission peak, as demonstrated through RK45-based ion trajectory simulation and SIMION.

A detailed assessment of QMF performance under radial asymmetry reveals that both mass resolution and transmission efficiency depend strongly on the asymmetry parameter and the polarity of the applied DC voltage. The maximum resolution is achieved for an outward displacement of the rod pair ($\gamma_x = +0.04$) when the displaced electrodes are biased at $+U$, while a comparable resolution is obtained for inward displacement ($\gamma_x = -0.04$) with reversed DC polarity. Although similar peak resolutions can be realized for both displacement configurations under appropriate polarity selection, outward displacement provides higher transmission acceptance due to the increased effective aperture of the mass filter. The pronounced dependence of resolution on DC polarity for a given asymmetry is explained by octupole-induced shifts of the stability boundary, confirmed through SIMION simulations.

The reduced transmission efficiency observed in asymmetric configurations is attributed to octupole-driven modification of RF phase selectivity at the QMF entrance, leading to a reduction in spatial acceptance. These findings demonstrate that controlled radial asymmetry, combined with appropriate DC polarity selection, offers a practical approach to tailoring resolution–transmission trade-offs in quadrupole mass filters operated in the second stability zone.

\begin{acknowledgement}
ND thanks SERB/ANRF India (CRG/2023/001529) and BRNS India (58/14/21/2023 - BRNS12329) for funding. AD acknowledges DST Inspire for research fellowship.
\end{acknowledgement}

\section{Conflicts of Interest}
The authors declare no conflicts of interest for this manuscript.



\bibliography{reference}

@article{dawson1984second,
  title={The second stability region of the quadrupole mass filter. II. Experimental results},
  author={Dawson, PH and Bingqi, Yu},
  journal={International journal of mass spectrometry and ion processes},
  volume={56},
  number={1},
  pages={41--50},
  year={1984},
  publisher={Elsevier}
}

@article{paul1955elektrische,
  title={Das elektrische massenfilter},
  author={Paul, Wolfgang and Raether, M},
  journal={Zeitschrift f{\"u}r Physik},
  volume={140},
  number={3},
  pages={262--273},
  year={1955},
  publisher={Springer}
}

@article{douglas2005linear,
  title={Linear ion traps in mass spectrometry},
  author={Douglas, Donald J and Frank, Aaron J and Mao, Dunmin},
  journal={Mass spectrometry reviews},
  volume={24},
  number={1},
  pages={1--29},
  year={2005},
  publisher={Wiley Online Library}
}

@article{moradian2007experimental,
  title={Experimental investigation of mass analysis using an island of stability with a quadrupole with 2.0\% added octopole field},
  author={Moradian, Annie and Douglas, DJ},
  journal={Rapid Communications in Mass Spectrometry: An International Journal Devoted to the Rapid Dissemination of Up-to-the-Minute Research in Mass Spectrometry},
  volume={21},
  number={20},
  pages={3306--3310},
  year={2007},
  publisher={Wiley Online Library}
}

@article{du1999peak,
  title={Peak splitting with a quadrupole mass filter operated in the second stability region},
  author={Du, Zhaohui and Douglas, DJ and Konenkov, Nikolai},
  journal={Journal of the American Society for Mass Spectrometry},
  volume={10},
  number={12},
  pages={1263--1270},
  year={1999},
  publisher={Springer}
}

@article{jana2025radial,
  title={Radial asymmetry in quadrupole mass filters: stability, multipole fields and resolution enhancement},
  author={Jana, Sukanya and Bose, Snigdha and Chakraborty, Sayel and Mandal, Pintu and Deb, Nabanita},
  journal={Journal of the American Society for Mass Spectrometry},
  year={2025},
  publisher={ACS Publications}
}

@article{ding2003quadrupole,
  title={Quadrupole mass filters with octopole fields},
  author={Ding, Chuanfan and Konenkov, NV and Douglas, DJ},
  journal={Rapid Communications in Mass Spectrometry},
  volume={17},
  number={22},
  pages={2495--2502},
  year={2003},
  publisher={Wiley Online Library}
}

@article{holme1973dependence,
  title={The dependence of ion transmission on the initial rf phase of a quadrupole mass filter},
  author={Holme, AE and Thatcher, WJ},
  journal={International Journal of Mass Spectrometry and Ion Physics},
  volume={10},
  number={3},
  pages={271--277},
  year={1973},
  publisher={Elsevier}
}

@article{hogan2008performance,
  title={Performance simulation of a quadrupole mass filter operating in the first and third stability zones},
  author={Hogan, Thomas J and Taylor, Stephen},
  journal={IEEE Transactions on Instrumentation and Measurement},
  volume={57},
  number={3},
  pages={498--508},
  year={2008},
  publisher={IEEE}
}

@article{miller1986quadrupole,
  title={The quadrupole mass filter: basic operating concepts},
  author={Miller, Philip E and Denton, M Bonner},
  journal={Journal of chemical education},
  volume={63},
  number={7},
  pages={617},
  year={1986},
  publisher={ACS Publications}
}

@article{li2021towards,
  title={Towards higher sensitivity of mass spectrometry: A perspective from the mass analyzers},
  author={Li, Chang and Chu, Shiying and Tan, Siyuan and Yin, Xinchi and Jiang, You and Dai, Xinhua and Gong, Xiaoyun and Fang, Xiang and Tian, Di},
  journal={Frontiers in chemistry},
  volume={9},
  pages={813359},
  year={2021},
  publisher={Frontiers Media SA}
}

@article{konenkov2002matrix,
  title={Matrix methods for the calculation of stability diagrams in quadrupole mass spectrometry},
  author={Konenkov, NV and Sudakov, M and Douglas, DJ},
  journal={Journal of the American Society for Mass Spectrometry},
  volume={13},
  number={6},
  pages={597--613},
  year={2002},
  publisher={Springer}
}

@article{ma1996simulation,
  title={Simulation of ion trajectories through the mass filter of a quadrupole mass spectrometer},
  author={Ma, FM and Taylor, S},
  journal={IEE proceedings. Science, measurement and technology},
  volume={143},
  number={1},
  pages={71--6},
  year={1996}
}

@article{konenkov1991characteristics,
  title={Characteristics of a quadrupole mass filter in the separation mode of a few stability regions},
  author={Konenkov, NV and Kratenko, VI},
  journal={International journal of mass spectrometry and ion processes},
  volume={108},
  number={2-3},
  pages={115--136},
  year={1991},
  publisher={Elsevier}
}

@article{douglas1992collisional,
  title={Collisional focusing effects in radio frequency quadrupoles},
  author={Douglas, DJ and French, John Barry},
  journal={Journal of the American Society for Mass Spectrometry},
  volume={3},
  number={4},
  pages={398--408},
  year={1992},
  publisher={Elsevier}
}

@article{dawson1975acceptance,
  title={The acceptance of the quadrupole mass filter},
  author={Dawson, PH},
  journal={International Journal of Mass Spectrometry and Ion Physics},
  volume={17},
  number={4},
  pages={423--445},
  year={1975},
  publisher={Elsevier}
}

@article{konenkov20174d,
  title={4D phase-space acceptance of the quadrupole mass filter},
  author={Konenkov, AN and Konenkov, NV and Verenchikov, AN},
  journal={European Journal of Mass Spectrometry},
  volume={23},
  number={3},
  pages={116--121},
  year={2017},
  publisher={SAGE Publications Sage UK: London, England}
}

@article{turner2005effect,
  title={Effect of ion entry acceptance conditions on the performance of a quadrupole mass spectrometer operated in upper and lower stability regions},
  author={Turner, P and Taylor, S and Gibson, JR},
  journal={Journal of Vacuum Science \& Technology A},
  volume={23},
  number={3},
  pages={480--487},
  year={2005},
  publisher={AIP Publishing}
}

@article{hogan2009effects,
  title={Effects of mechanical tolerances on QMF performance for operation in the third stability zone},
  author={Hogan, Thomas J and Taylor, Stephen},
  journal={IEEE Transactions on Instrumentation and Measurement},
  volume={59},
  number={7},
  pages={1933--1940},
  year={2009},
  publisher={IEEE}
}

@article{schulte1999nonlinear,
  title={Nonlinear field effects in quadrupole mass filters},
  author={Schulte, J and Shevchenko, PV and Radchik, AV},
  journal={Review of scientific instruments},
  volume={70},
  number={9},
  pages={3566--3571},
  year={1999},
  publisher={American Institute of Physics}
}

@article{bugrov2023modelling,
  title={Modelling some rod set imperfections of a quadrupole mass filter},
  author={Bugrov, Pavel V and Sysoev, Aleksey A and Konenkov, Nikolai V},
  journal={Journal of Mass Spectrometry},
  volume={58},
  number={12},
  pages={e4986},
  year={2023},
  publisher={Wiley Online Library}
}

@book{dawson2013quadrupole,
  title={Quadrupole mass spectrometry and its applications},
  author={Dawson, Peter H},
  year={2013},
  publisher={Elsevier}
}

@book{march2005quadrupole,
  title={Quadrupole ion trap mass spectrometry},
  author={March, Raymond E and Todd, John F},
  year={2005},
  publisher={John Wiley \& Sons}
}

@article{JANA2025117495,
title = {Effect of rod asymmetry on field distortions and transmission characteristics in linear quadrupole mass filters},
author = {Sukanya Jana and Santu Nath and Pintu Mandal and Nabanita Deb},
journal = {International Journal of Mass Spectrometry},
volume = {517},
pages = {117495},
year = {2025},
issn = {1387-3806},}

@article{paul1953neues,
  title={Ein neues massenspektrometer ohne magnetfeld},
  author={Paul, Wolfgang and Steinwedel, Helmut},
  journal={Zeitschrift f{\"u}r Naturforschung A},
  volume={8},
  number={7},
  pages={448--450},
  year={1953},
  publisher={Verlag der Zeitschrift f{\"u}r Naturforschung}
}

@article{dahl2000simion,
  title={SIMION 3D Version 7.0 User’s manual},
  author={Dahl, David A}
}

\end{document}